\documentclass[prl, aps, twocolumn, amsmath, amssymb, superscriptaddress, nolongbibliography]{revtex4-2}
\usepackage{graphicx,verbatim}
\usepackage{braket}
\usepackage{color}
\usepackage[colorlinks=true,urlcolor=black]{hyperref}

\usepackage{soul}
\usepackage[normalem]{ulem}
\usepackage[x11names]{xcolor}

\begin{document}

\title{Numerically exact simulation of photo-doped Mott insulators}

\author{Fabian Künzel}
\affiliation{
Institute of Theoretical Physics, University of Hamburg, 20355 Hamburg, Germany
	}
\author{André Erpenbeck}
\affiliation{
	Department of Physics, University of Michigan, Ann Arbor, Michigan 48109, USA
	}
\author{Daniel Werner}
\affiliation{
Institute of Theoretical and Computational Physics, Graz University of Technology, 8010 Graz, Austria
	}
\author{Enrico Arrigoni}
\affiliation{
Institute of Theoretical and Computational Physics, Graz University of Technology, 8010 Graz, Austria
	}	
	\author{Emanuel Gull}
\affiliation{
	Department of Physics, University of Michigan, Ann Arbor, Michigan 48109, USA
	}
	\author{Guy Cohen}
\affiliation{
The Raymond and Beverley Sackler Center for Computational Molecular and Materials Science, Tel Aviv University, Tel Aviv 6997801, Israel
	}
\affiliation{
School of Chemistry, Tel Aviv University, Tel Aviv 6997801, Israel
	}
\author{Martin Eckstein}
\affiliation{
Institute of Theoretical Physics, University of Hamburg, 20355 Hamburg, Germany
	}	
\affiliation{	
	The Hamburg Centre for Ultrafast Imaging, Hamburg, Germany
	}
	
\date{\today}

\begin{abstract}
A description of long-lived photo-doped states in Mott insulators is challenging, as it needs to address exponentially separated timescales. We demonstrate how properties of such states can be computed using numerically exact steady state techniques, in particular Quantum Monte Carlo, by using a time-local ansatz for the distribution function with separate Fermi functions for the electron and hole quasiparticles. The simulations show that the Mott gap remains robust to large photo-doping, and the photo-doped state has hole and electron quasiparticles with strongly renormalized properties. 
\end{abstract}

\maketitle

\paragraph{Introduction ---}
Short light pulses provide intriguing avenues to manipulate material properties on ultrafast timescales \cite{Basov2017, Giannetti2016, Sentef2021}.  Mott insulators are particularly interesting in this regard \cite{Murakami2023}, because a zoo of complex orders can emerge from a perturbed Mott phase. A versatile route toward the generation of non-thermal phases involves photo-doping, i.e., the creation of charge carriers such as doublons (doubly occupied sites)  and holes in a single-orbital Mott insulator. With increasing gap, carrier recombination becomes exponentially slow \cite{Rosch2008a,Sensarma2010a,Zala2013PRL}, so that doublon and hole densities are approximately conserved over extended periods. Energy dissipation into the spin and phonon background eventually yields a cold state akin to electron-hole liquids in semiconductors \cite{Keldysh1986review}. Such cold photo-doped phases in correlated electron systems may undergo metal insulator transitions and band reconstruction \cite{Sandri2015,he2016,Wegkamp2014,lantz2017,verma2023,Beaulieu2021}, and potentially even manifest superconducting instabilities \cite{Li_Eta_2020,Ray2023,Murakami2022,Werner2019b}.

Dynamical Mean Field Theory (DMFT) \cite{Georges_Dynamical_1996} and its extensions \cite{Maier2005, Rohringer2018} are a powerful approach to 
study 
Mott materials. 
The main challenge in extending these methods to the time domain is the solution of a  quantum impurity model. Real-time nonequilibrium DMFT simulations  \cite{Aoki_Nonequilibrium_2014,Freericks_Nonequilibrium_2006} based on 
numerically exact Quantum Monte Carlo (QMC) \cite{Muehlbacher2008,Werner2009,Eckstein2009} or matrix-product states \cite{Wolf2014,Balzer2015} have been limited to short times, hindering the study of cold photo-doped states. Presently, 
state-of-the-art methods to study photo-doped Mott insulators are 
perturbative variants of the strong-coupling expansion \cite{Keiter_Perturbation_1970, Coleman_New_1984, Eckstein2010nca}, notably the Non-Crossing Approximation (NCA),
which unfortunately is least reliable in the most relevant metallic regime.
Conversely, significant advance has been made with non-perturbative  techniques aimed at the nonequilibrium steady state,
through the auxiliary master equation formalism (AMEA) \cite{Arrigoni2013}, and more recently the steady state variant 
 \cite{Erpenbeck_Quantum_2023} of the inchworm algorithm \cite{Cohen_Taming_2015}, a high-order stochastic evaluation of the self-consistent strong-coupling expansion.

In this letter, we aim to use such potentially 
numerically exact 
steady state solvers to investigate slowly evolving (quasi-steady)  photo-doped states. Previously, 
quasi-steady photo-doped states have 
been modeled as an equilibrium state of an approximate large-$U$ Hamiltonian which {\em exactly} conserves the doublon and hole densities \cite{Murakami2022}, and, following ideas introduced in \cite{lange2017}, by maintaining the nearly conserved  doublon and hole densities by 
external charge reservoirs \cite{Li_Nonequilibrium_2021,atanasova_correlated_2020}.  Here we introduce an approach that is not 
restricted 
to large U and does not alter the system by additional reservoirs. As in quantum kinetic equations \cite{Picano2021,KamenevBook}, we take the distribution function $F(\omega,t)$ to be a dynamical variable.
 The non-perturbative steady state solvers mentioned above can be used to solve the many-body problem with any given distribution function $F_{\rm steady}(\omega)$. The equilibrium state is a special case where fluctuation-dissipation relations guarantee that $F_{\rm steady}(\omega)$  is equal to the Fermi function $f(\omega)$. 
 One can therefore introduce a time-local-$F$ ansatz (TLFA), taking  the steady state solution with $F(\omega,t)=F_{\rm steady}(\omega)$ as an approximate description of  the slowly evolving state around time $t$. Below, we validate the accuracy of the TLFA through real-time simulations, and use the ansatz to simulate complex photo-doped phases with non-perturbative techniques. 

\paragraph{Model ---}
We consider the one-band Hubbard model at half filling, with Hamiltonian 
\begin{align}
H = -\tilde t_0 \sum_{\sigma\langle i,j \rangle} c_{\sigma i}^\dagger c_{\sigma j} + U \sum_{i} n_{i\uparrow}n_{i\downarrow} -\frac{U}{2}\sum_{\sigma i} n_{i\sigma}.
\label{eq:hamiltonian}
\end{align}
Here, $c_{\sigma i}^{(\dagger)}$ are electronic annihilation (creation) operators at site $i$ and spin $\sigma$, $n_{i\sigma}=c_{\sigma i}^{\dagger}c_{\sigma i}$, $U$ is the Coulomb repulsion, and $\tilde t_0$ the hopping amplitude between nearest-neighbor sites $\langle i,j\rangle$. We solve this system by means of nonequilibrium DMFT
\cite{Aoki_Nonequilibrium_2014} 
on the  Bethe lattice with coordination number $z\to\infty$ and hopping $\tilde t_0=t_0/\sqrt{z}$. We use a hopping $t_0=\sqrt{2}$ (bandwidth $8$) and $\hbar=1$, i.e., all energies are measured in $t_0/\sqrt{2}$ and times are measured in units of $\sqrt{2}/t_0$.
To describe the energy dissipation, we include an external bosonic bath via a phonon self-energy $\Sigma_{\mathrm{ph}}(t,t') = g^2G(t,t')D_{\mathrm{bath}}(t,t'),$ with coupling strength $g^2=0.5$ and 
a linear density of states $D_{\mathrm{bath}} (\omega) = \frac{\omega}{\omega_c^2}e^{-\frac{\omega}{\omega_c}}$ \cite{Eckstein_Photoinduced_2013,peronaci2018}. The cutoff $\omega_c = 0.2 \ll U$ is chosen such that only kinetic energy relaxation of doublons and holes is possible, while direct recombination via phonon emission is not.

 \begin{figure}[tbp]
\centerline{\includegraphics[width= 1.0\columnwidth]{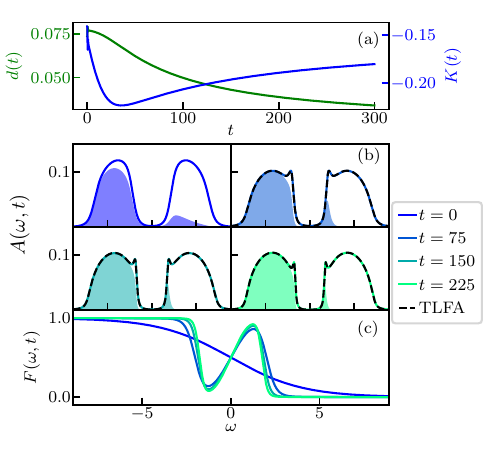}}
\caption{Time evolution of the photoexcited state, initially prepared at $T_i=2.0$. 
(a) Expectation value of the double occupancy $d(t)$ (left axis) an kinetic energy $K(t)$ (right axis).
(b) Spectral function $A(\omega,t)$ (solid line) and occupied density of states $A^<(\omega,t)$ (shaded area) for different representative times. (c) Distribution function $F(\omega,t)$ at different times.
Dashed lines in (b) show the spectra obtained from the TLFA, with the corresponding distribution function taken from (c).
}
\label{fig:real-time}
\end{figure} 

{\em TLFA ---}
For validating the 
TLFA, 
we solve the model in Eq.~\eqref{eq:hamiltonian} in real-time in a setting similar to Ref.~\cite{Nagamalleswararao_Photoinduced_2021}, using DMFT + NCA within the NESSi simulation package \cite{Nessi}. We start from the Mott phase ($U=8$) at a high initial temperature  $T_i=2$  and monitor the evolution as energy is dissipated to the bosonic  bath at lower temperature 
$T_{\rm bath}={1/12.5}$. 
Such a temperature quench is convenient to initiate the dynamics with a given density of doublons and holes, but an analogous long-time dynamics is expected if the initial state population is generated by a short pulse \cite{Nagamalleswararao_Photoinduced_2021}. The doublon-hole recombination after the quench is evident in Fig.~\ref{fig:real-time}(a) through the  slow decay of the double occupancy $d(t) = \langle n_{\uparrow}(t)n_{\downarrow}(t) \rangle $.
 The kinetic energy $K(t)$ shows 
a much faster initial drop, which reflects the initial intra-band relaxation process due to the phonon bath and a slower increase associated to doublon-holon recombination.
This temporal separation 
is more pronounced for larger gaps
\cite{supplement}.

To analyze spectral and distribution functions, we perform a partial Fourier transformation (Wigner transform) of the real-time Green's functions  $G^{R,<}(\omega,t)=\int ds\, e^{i\omega s} G^{R,<}(t+s/2,t-s/2)$  at average time $t$, using a fixed window $|s|\le 150$ for the relative time $s$. The spectral function and distribution function are then given by $A(\omega,t)=- \frac 1 \pi \mathrm{Im} \{G^R(\omega,t) \} $ and $F(\omega,t) = -\frac 1 2 \mathrm{Im}\{G^<(\omega,t)\}/\mathrm{Im}\{G^R(\omega,t)\}$, respectively. The spectral function [Fig.~\ref{fig:real-time}(b)] starts from two Hubbard bands with a partially filled gap due to the high initial temperature. As the kinetic energy of the doublons relaxes within a few tens of inverse hoppings, the occupied density of states $A^<(\omega,t) \equiv F(\omega,t)A(\omega,t)$ concentrates at the lower band edge of the upper Hubbard band. At the same time, two peaks emerge in the spectrum at the edges of the Mott gap, which indicate the simultaneous presence of doublon and hole  quasiparticles \cite{Eckstein_Photoinduced_2013, Werner2019b, Nagamalleswararao_Photoinduced_2021}. Correspondingly, the distribution function develops two separate quasiparticle chemical potentials for the hole and doublon charge carriers [Fig.~\ref{fig:real-time}(c)]. These  spectral characteristics slowly relax back to equilibrium as doublons and holes recombine. 

To implement the TLFA, we extract the function $F(\omega,t)$ at a particular time $t$ and determine a nonequilibrium steady state solution with distribution function $F_{\rm steady}(\omega)=F(\omega,t)$. In practice, we solve the DMFT impurity model with a time-translationally invariant hybridization function whose spectral (retarded) component $\Delta^R(\omega)$ is determined through the DMFT self-consistency, while the lesser component is determined by the given distribution function $F_{\rm steady}(\omega)$, i.e.,  $\Delta^<(\omega)=-\frac{1}{2} F_{\rm steady}(\omega) \text{Im} \Delta^R(\omega)$. The resulting spectral functions $A_{\rm TLFA}(\omega)$ are shown by dashed lines in Fig.~\ref{fig:real-time}(b). They almost perfectly reproduce the real-time spectra, 
i.e., the distribution function $F(\omega,t)$ characterizes the system at time $t$, without further dependence on the history.

\paragraph{Nonequilibrium steady state spectral functions in the photo-doped 
system ---}
The above validation as well as previous studies \cite{Nagamalleswararao_Photoinduced_2021} motivate us to 
use the TLFA to obtain the spectral function of photo-doped systems using the numerically exact steady state inchworm algorithm. 
Fig.~\ref{fig:real-time}(c) suggests to adopt an
 ansatz for the distribution function, which interpolates between Fermi functions $f(\omega\mp\mu_{\rm ex}, T)$ with generalized chemical potentials $\pm\mu_{\rm ex}$ for the electron-like ($\omega>0$) and hole-like ($\omega<0$) side, respectively. More precisely,  $F_{T,\mu_{\rm ex}}(\omega) = \Theta_\alpha(\omega) f(\omega+ \mu_{\rm ex}, T)  + (1-\Theta_\alpha(\omega)) f(\omega-\mu_{\rm ex}, T)$, with a smooth step-function  $\Theta_\alpha(\omega) = 0.5\bigl(1-\tanh(\omega\alpha/2)\bigr)$.
The interpolation affects $F_{T,\mu_{\rm ex}}(\omega)$ only 
within 
the gap, so that results are largely independent of the parameter $\alpha$ 
\cite{supplement}; 
below we choose $\alpha=\beta$. 
We then 
fix a given  photo-doping density,
\begin{align}
n_{\rm ex}(T,\mu_{\rm ex}) =-\frac{1}{\pi}
\int_0^{\infty}d\omega\, F_{T,\mu_{\rm ex}}(\omega) \text{Im}\{\Delta^R(\omega)\},
\label{nx}
\end{align}
by adapting  $\mu_{\rm ex}$. 
At half filling, 
spectra and quasiparticle properties are symmetric with respect to hole and doublon excitations.

\begin{figure}[tbp]
\centerline{\includegraphics[width=1.0\columnwidth]{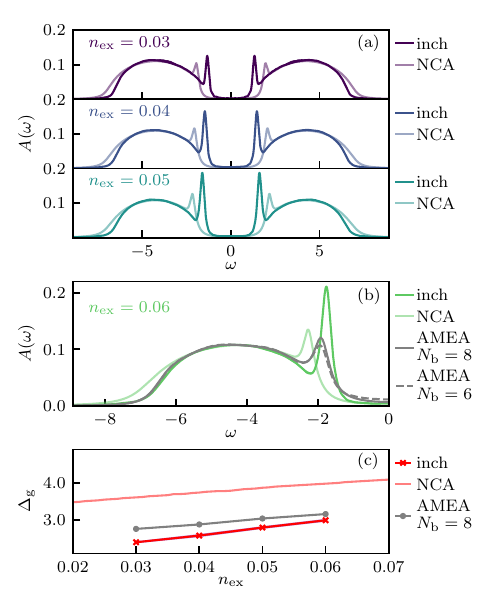}}
\caption{(a) Comparison of the spectral function $A(\omega)$ for given $n_{\rm ex}$, obtained by inchworm QMC (solid lines) and by NCA (transparent lines), at temperature $T_{\Delta}=1/12.5$. (b) Comparison as in (a) with AMEA spectra using 6 (dashed gray line) and 8 bath sites (solid grey line), for $n_{\rm ex}=0.06$.  (c) The Mott gap $\Delta_{\rm g}$ for all three approaches. The gap is defined by the spectral weight reaching 0.057, which is half of the maximum of the equilibrium spectrum. For a visualization of the Monte Carlo error, the mean of five inchworm DMFT iterations is plotted together with the standard deviation (shaded region).
}
\label{fig:spectra}
\end{figure} 

The inchworm algorithm computes the time-translationally invariant Green's functions $G^{R,<}(t-t')$ from the real-time hybridization function  $\Delta^{R,<}(t-t')$. In each DMFT iteration, we transform $G^R(t-t')$ to obtain $G^R(\omega)$, determine the self-consistent  $\Delta^R(\omega)$, set $\Delta^<(\omega)=-\frac{1}{2}\text{Im}\Delta(\omega) F_{T,\mu_{\rm ex}}(\omega)$, where $\mu_{\rm ex}$ is determined to match condition \eqref{nx} for a given $n_{\rm ex}$, obtain $\Delta^R(t)$ and $\Delta^<(t)$ from the inverse Fourier transform, and perform the Monte Carlo evaluation of $G$. For details of the Monte Carlo implementation we refer to \cite{Erpenbeck_Quantum_2023}. The convergence of the QMC data with the DMFT iteration and with the diagrammatic order is analyzed in the supplemental material \cite{supplement}. We note that the DMFT iteration based on the TLFA is easier to converge compared to a conventional steady state setup, where the solution depends on external reservoirs \cite{Li_Nonequilibrium_2021,atanasova_correlated_2020}. In the latter case, both  $\Delta^R(\omega)$ and $\Delta^<(\omega)$ would be determined from independent self-consistency conditions, and the additional Monte Carlo noise in $\Delta^<(\omega)$ slows convergence.

In Fig.~\ref{fig:spectra}(a), we compare the spectral function $A(\omega)$ obtained for NCA and the numerically exact inchworm QMC. The main characteristics of the cold photo-doped state, which is the simultaneous hole and electron quasiparticle peak, is thereby validated by the numerically exact data. From the spectra we can extract the gap $\Delta_{\rm g}$, see Fig.~\ref{fig:spectra}(c). Here, NCA overestimates the gap $\Delta_{\rm g}$ in the photo-doped state, consistent with its behavior in 
equilibrium 
\cite{pruschke1989anderson}.
Suppressing higher-order diagrams essentially
increases the effective interaction strength and leads to 
a 
larger gap. 
Nevertheless, one finds that the gap remains robust at large photo-doping $n_{\rm ex}$ even in the numerically exact solution, which is an important finding supporting the stability of photo-doped orders.

 \begin{figure}[tbp]
\centerline{\includegraphics[width=1.0\columnwidth]{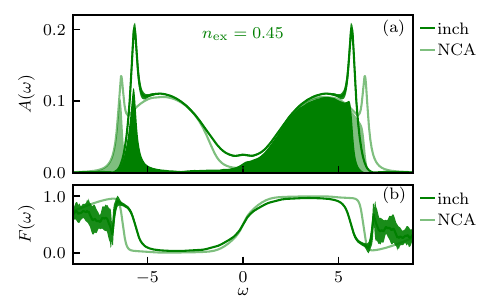}}
\caption{(a) Spectral function $A(\omega)$ at $n_{\rm ex}=0.45$, obtained with inchworm QMC (solid lines) and NCA (transparent lines), at temperature $T_{\Delta}=1/12.5$. The colored areas denote the occupied spectrum $A^<(\omega)$. (b) Comparison of the inchworm QMC (solid lines) and NCA (transparent lines) distribution function $F(\omega)$ at $n_{\rm ex}=0.45$ and temperature $T_{\Delta}=1/12.5$. The mean of five inchworm DMFT iterations is plotted together with the standard deviation (shaded region).
}
\label{fig:large_ex}
\end{figure} 

We also compute NCA and inchworm QMC spectra for a photo-doping close to population inversion, see Fig.~\ref{fig:large_ex}(a). Also in this extreme case, the inchworm code validates the photo-doped state and shows characteristic quasi-particle peaks at the outer edges of the Hubbard band as well as a superposition of two separate Fermi functions in the distribution function $F(\omega)$ in Fig.~\ref{fig:large_ex}(b). This supports the stability of states with large photo-doping, which have also been observed 
in DMFT+NCA simulations of the Hubbard model  using other doping protocols \cite{Werner2019c,Li_Eta_2020}.

The TLFA can also be evaluated with the AMEA, which approximates the impurity problem with hybridization function  $\Delta^R(\omega)$ and $\Delta^<(\omega)$ in terms of a finite open system described by $N_{\rm b}$ bath orbitals with additional Lindblad dissipators. For a detailed description of the bath fitting procedure, see Ref.~\cite{Dorda_Auxiliary_2014, Dorda_Optimized_2017}. 
Relatively inexpensive simulations are possible with up to $N_{\rm b}=8$ sites within a configuration interaction expansion~\cite{Werner_Configuration_2023}. While one can see in Fig.~\ref{fig:spectra}(b) that these data are not yet converged as a function of $N_{\rm b}$, the difference between $N_{\rm b}=6$ to $8$ sites indicates  the correct trend. Even simulations with only $N_{\rm b}=6$ sites provide a significant improvement over the NCA simulation regarding the size of the gap, and accurately capture the high-energy behavior of the spectra.
 
 \begin{figure}[tbp]
\centerline{\includegraphics[width=1.0\columnwidth]{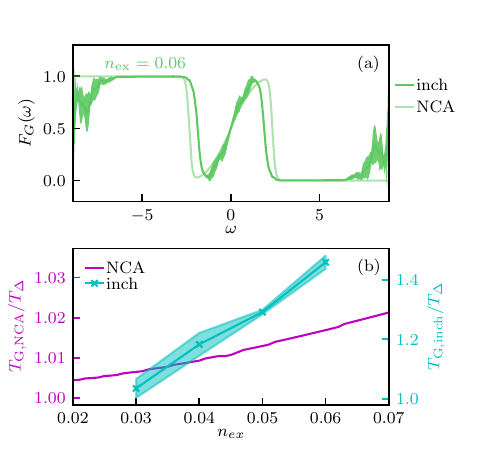}}
\caption{(a) 
Distribution 
function $F(\omega)$ at $n_{\rm ex}=0.06$, obtained with inchworm QMC (solid lines) and NCA (transparent lines), at temperature $T_{\Delta}=1/12.5$.  (b) Effective temperature $T_{\rm G,NCA}/T_{\Delta}$ (left axis) and $T_{\rm G, inch}/T_{\Delta}$ (right axis) of the quasi-particle excitations using the two methods in relation to $T_{\Delta}$. For a visualization of the Monte-Carlo error, the mean of five inchworm DMFT iterations is plotted together with the standard deviation (shaded region).
}
\label{fig:gap}
\end{figure} 

For the results presented here, we used an universal ansatz for $F_{\rm steady}(\omega)$, whose form is motivated by previous real-time simulations. 
An interesting question is if one can validate that this ansatz corresponds to an actual long lived state. To answer this question, one could compute the time-evolution of this state using quantum Boltzmann equations (QBE) \cite{Picano2021}. The latter can be formulated as 
an evolution equation $\partial_t F(\omega,t) = I[F(\omega,t)]$ for $F$,  
where the scattering  integral $I[F]$ is determined in terms of the self-energy $\Sigma_{\rm TLFA}[F]$ obtained from a TLFA. Solving the QBE with AMEA or inchworm QMC is beyond the scope of the present work. However, we note that an infinitely long-lived state would correspond to a case in which the distribution functions of all quantities coincide, i.e., $F_G=F_\Delta=F_\Sigma$, where $F_{X}=-\tfrac12\text{Im} X^R(\omega)/\text{Im} X^<(\omega)$. Because $F_\Delta$ is prescribed in the present implementation of the TLFA, a useful quantity to analyze is the difference between $F_\Delta$ and $F_G$, which is obtained from the computed Green's function. In Fig.~\ref{fig:gap}(a), we show an example of the distribution $F_G$. To quantify how it differs from $F_\Delta$, we extract a temperature $T_{G}$ from linear fits of the form $-(\omega-\mu_{\rm ex,G})/T_G$ to the function $\log\bigl\{F_G(\omega)/\bigl(1-F_G(\omega)\bigr)\bigr\}$ in the vicinity of the quasiparticle peak. Figure \ref{fig:gap}(b) shows that $F_G$ is consistently larger than $F_\Delta$. The difference is larger in the inchworm results, which is expected as the smaller gap would imply a faster dynamics,
so that the state found by the TLFA has a shorter lifetime and eventually thermalizes.

\paragraph{Conclusion ---}
Optically excited Mott insulators exhibit slowly evolving quasi-steady photo-doped states, that are challenging to describe theoretically.
In this work, we have demonstrated 
how properties of these long-lived photo-doped states can be accessed with numerically exact techniques, by using a time-local ansatz for the electronic distribution function.
We have validated the consistency of this ansatz upon comparison with real-time simulations in a 
quenched Hubbard model. 
Employing a universal form of the distribution function
we directly calculate photo-doped Mott spectra for various photo-excitation levels using steady state NCA, inchworm QMC, and AMEA. 
The resulting photo-doped spectra 
can be converged
in a
wide
range of doping densities, and the Mott gap remains robust up to large photo-dopings for all of the methods. 
While NCA 
overestimates 
the gap, AMEA shows a trend towards the direction of the numerically exact inchworm QMC method.
These findings indicate that nonequilibrium steady state formalisms can be used to directly access quasi-stable photo-doped states in Mott insulators. 
Furthermore, by integrating the steady state ansatz with QBE schemes, they open up new avenues for characterizing the slow dynamics of Mott insulators.
This approach has the potential to extend into time scales far beyond the capabilities of existing real-time simulations. 
\\

F.K. and M.E. were funded by the Deutsche Forschungsgemeinschaft through 
QUAST-FOR5249 -  449872909 (Project P6).
Until August 31, A.E.~was funded by the Deutsche Forschungsgemeinschaft -- 453644843.
E.G.~and A.E., starting on September 1, were supported by the U.S. Department of Energy, Office of Science, Office of Advanced Scientific Computing Research and Office of Basic Energy Sciences, Scientific Discovery through Advanced Computing (SciDAC) program under Award Number DE‐SC0022088.
G.C.~acknowledges support by the Israel Science Foundation (Grants No.~2902/21 and 218/19) and by the PAZY foundation (Grant No.~318/78).
E.A. and D.W. acknowledge funding by the Austrian Science Fund 
(FWF, Grant No. P 33165-N) 
and by NaWi Graz. The computations were performed on the HPC-Cluster of the PHYSnet-Rechenzentrum University of Hamburg and the Vienna Scientific Cluster.
This research used resources of the National Energy Research Scientific Computing Center, a DOE Office of Science User Facility supported by the Office of Science of the U.S. Department of Energy under Contract No. DE-AC02-05CH11231 using NERSC award BES-ERCAP0021805. 
We thank Jan Lotze, Jiajun Li, and Philipp Werner for useful discussions. An early version of the non-equilibrium steady state code has been written by Jiajun Li.

\appendix

\clearpage

\section{Supplemental material for ``Numerically exact simulation of photo-doped Mott insulators''}
\subsection{Nonequilibrium Dynamical Mean Field Theory}
Our numerical calculations are based on nonequilibrium DMFT, which is a generalization of conventional DMFT for Green's functions defined on the L-shaped Kadanoff-Baym contour $\mathcal C$ \cite{Aoki_Nonequilibrium_2014}. The key idea is to assume the self-energy to be local and self-consistently map the lattice problem onto a single-impurity problem with action $\mathcal S$, which is given at half filling by
\begin{align}
\mathcal S=& -i\int_{\mathcal C} dt \,U(t) \big(n_{\uparrow}(t)-\tfrac 1 2\big) \big(n_{\downarrow}(t)-\tfrac 1 2\big) \nonumber \\
&-i \sum_{\sigma=\uparrow , \downarrow} \int_{\mathcal C} dt dt' c_{\sigma}^\dagger(t) \Delta(t,t') c_{\sigma}(t'),
\end{align}
with $\Delta(t,t')$ being the hybridization function. 
Coupling the system to a phononic bath will add a contribution to the self-energy $\Sigma(t,t')$,
\begin{align}
\Sigma(t,t')= \Sigma_{\mathrm U}(t,t')+\Sigma_{\mathrm{ph}}(t,t'),
\end{align}
which now consists of a phononic ($\Sigma_{\mathrm{ph}}$) and lattice ($\Sigma_{\mathrm U}$) part. Together with the noninteracting Green's function $\mathcal G(t,t')$ and the Dyson equation for the impurity Green's function $G_{\mathrm{imp}}(t,t')$
\begin{align}
\mathcal G(t,t')^{-1}&= [i \partial_t + \mu - h(t)]\delta_{\mathcal C}(t,t')-\Delta(t,t'), \nonumber \\
G_{\mathrm{imp}}(t,t')^{-1} &= \mathcal G(t,t')^{-1} -\Sigma(t,t'),
\end{align}
one can see, that the phononic bath $\Sigma_{\mathrm{ph}}(t,t')$ can be equivalently added to the hybridization $\Delta(t,t')$ instead of $\Sigma(t,t')$.
In our case, i.e. given a semi-elliptic density of states, this leads to the following Bethe lattice self-consistency condition
\begin{align}
\Delta(t,t')=t_0^2 G_{\mathrm{loc}}(t,t')+\Sigma_{\mathrm{ph}}(t,t'),
\end{align}
with $G_{\mathrm{loc}}(t,t')$ being the local Green's function of the central Bethe lattice site.
Following previous studies of photo-doped Mott insulators \cite{Li_Nonequilibrium_2021,Nagamalleswararao_Photoinduced_2021}, we incorporate the phononic self-energy from the main text as an addition to the lattice self-consistent part of the hybridization, given by the local Green's function. In the steady state all Green's functions and hybridization functions are time translationally invariant and thus can be Fourier transformed into frequency domain.
\\

\subsection{Ansatz for the distribution function}
\begin{figure}[h!]
     \includegraphics{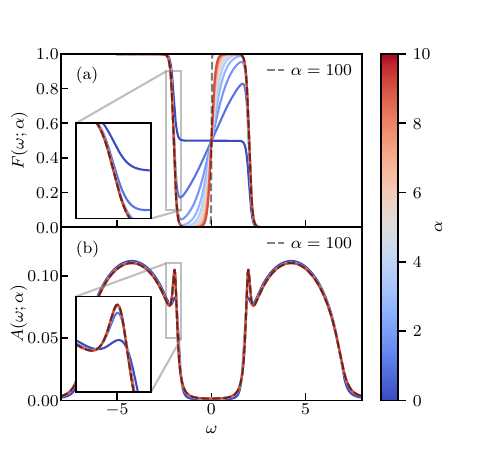}
    \caption{ 
                Dependency of the distribution function (a) and the resulting spectrum (b) on the parameter $\alpha$ used in the TLFA function $F_{T,\mu_{\rm ex}}(\omega)$ for $T=1/12.5$ and $\mu_{\rm {ex}}=2.0$ ($n_{\rm ex}=0.03$).
            }
    \label{fig:smooth}
\end{figure}
In the main manuscript, we propose an ansatz for the distribution function for the slowly evolving quasi-steady photo-doped states in Mott insulators of the form
$F_{T,\mu_{\rm ex}}(\omega) = \Theta_\alpha(\omega) f(\omega+ \mu_{\rm ex}, T)  + (1-\Theta_\alpha(\omega)) f(\omega-\mu_{\rm ex}, T)$,
where the smooth step-function  $\Theta_\alpha(\omega) = 0.5\bigl(1-\tanh(\omega\alpha/2)\bigr)$ interpolates between different Fermi distributions in the upper and lower Hubbard band. 
This function is motivated by the real-time data in Fig.~1 of the main text and by previous studies \cite{Nagamalleswararao_Photoinduced_2021}.
While the Fermi-functions for the electron- and hole-excitations are intuitive, 
the interpolation function $\Theta_\alpha(\omega)$, 
and especially the parameter $\alpha$, are somewhat arbitrary. 
In Fig.~\ref{fig:smooth}, we demonstrate that the results for the spectra are insensitive to the precise value of $\alpha$ across a  wide range of values. 
Specifically, the influence of $\alpha$ on the spectral function, as shown in Fig.~\ref
{fig:smooth}(b), is negligible for values of $\alpha \gtrsim 2$.
We explain this insensitivity with the fact that $\alpha$ only influences the distribution function in the region of the gap, where the spectral weight is insignificant. This observation, together with our effort to restrict the number of parameters of the ansatz, justifies our choice of $\alpha=\beta$. 
\\

\subsection{Large gap Mott systems}
\begin{figure}[h!]
     \includegraphics{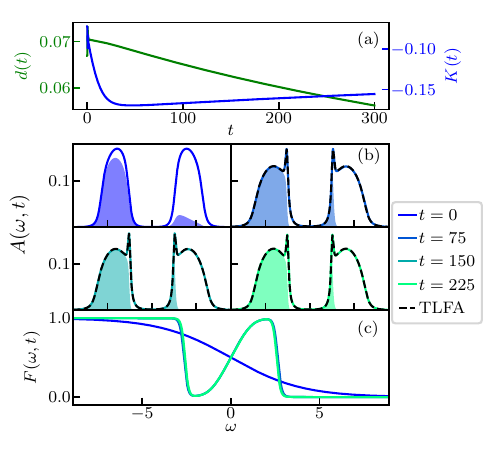}
    \caption{
               Time evolution of the photoexcited state, initially prepared at $T_i=2.0$, for a system with larger Mott gap ($U=8$, $t_0=1$). 
(a) Expectation value of the double occupancy $d(t)$ (left axis) an kinetic energy $K(t)$ (right axis).
(b) Spectral function $A(\omega,t)$ (solid line) and occupied density of states $A^<(\omega,t)$ (shaded area) for different representative times. (c) Distribution function $F(\omega,t)$ at different times.
Dashed lines in (b) show the spectra obtained from the TLFA, with the corresponding distribution function taken from (c).
            }
    \label{fig:large_gap}
\end{figure}

The main text states that the real-time dynamics of systems with larger Mott gaps exhibits a more pronounced separation of timescales. In order to demonstrate this relation, we calculate the real-time evolution of a system with larger 
gap, using $U=8$ and $t_0=1$. Fig.~\ref{fig:large_gap}(a)
shows that the initial drop of the kinetic energy $K(t)$ due to the intra-band relaxation happens on 
a similar timescale, regardless of the gap size. 
The subsequent increase of $K(t)$ due to recombination, however, takes place on a much longer timescale than for the system with smaller gap. This is expected, since the size of the gap affects the timescale of the charge carrier recombination, while for initial intra-band thermalization it is given by the phonon bath. 
Correspondingly, one observes a 
slower drop off of the double occupancy $d(t)$ (Fig.~\ref{fig:large_gap}(a)) and 
a
slower dynamics of the quasiparticle 
excitation. The 
spectra (Fig.~\ref{fig:large_gap}(b)) and the distribution function (Fig.~\ref{fig:large_gap}(c)) only evolve slowly after the initial thermalization. 
As for the smaller gap Mott insulators discussed in the main text, we find that  the TLFA reproduces the real-time spectra almost perfectly (dashed lines in Fig.~\ref{fig:large_gap}(b)).

\subsection{Convergence of the inchworm QMC results}

In the main manuscript, DMFT results calculated by the inchworm QMC method are discussed.
In order for this data to be reliable, the results need to be converged with respect to: (i) the DMFT self-consistency and (ii) the hybridization order considered for the inchworm scheme. Both these convergences are discussed in the following.

\begin{figure}[h!]
    \raggedright (a)\\ \centering
    \includegraphics{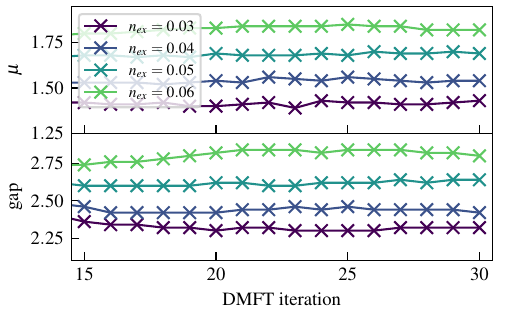}\\
    \raggedright (b)\\ \centering
    \includegraphics{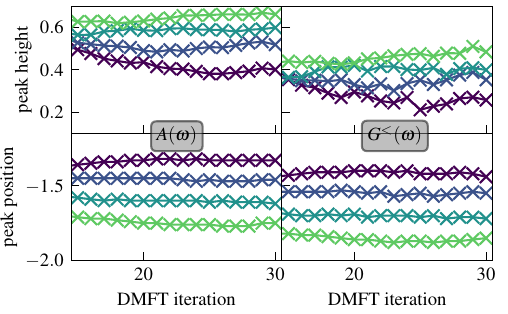}
    \caption{ 
            Convergence of the inchworm QMC method with respect to DMFT iterations. 
            This data pertains Fig.~2 of the main manuscript.
            (a) Chemical potential and gap size as a function of DMFT iteration for different photo-doping levels.
            (b) Behavior of the quasiparticle peaks, namely peak height and position, as a function of DMFT iteration for different photo-doping levels.
            }
    \label{fig:convergence_DMFT}
\end{figure}

Fig.~\ref{fig:convergence_DMFT} 
analyzes the behavior of different physically relevant parameters as a function of DMFT iterations.
Fig.~\ref{fig:convergence_DMFT}(a) 
shows that the chemical potential $\mu$, which is used in the TLFA to support a given photo-doping level, as well as the gap in the Mott system are stable throughout consecutive DMFT iterations. The minor variations are attributed to the Monte Carlo noise of the inchworm QMC scheme.
Fig.~\ref{fig:convergence_DMFT}(b) plots the behavior of different metrics for the quasiparticle resonances as a function of DMFT iterations.
In particular, the position and the height of the photo-excited quasiparticle is investigated (top and bottom panels), both for the spectrum and the lesser component of the GF (left and right panels).
While the position of the peaks is rather stable, we find some variations in the peak height, especially for the lesser GF. As these quantities are more susceptible to MC noise, and as we  do not observe any trends with increasing DMFT iteration count, we are confident that our results are converged with respect to the DMFT cycle. In order to minimize the influence of MC noise on the final result, we consider the last 5 DMFT iterations in the main manuscript.

\begin{figure}[h!]
    \raggedright (a)\\ \centering
    \includegraphics[width=0.9\columnwidth]{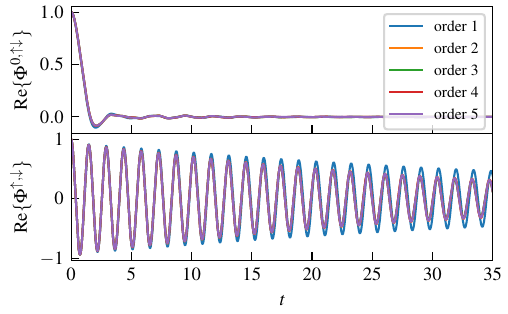}\\
    \raggedright (b)\\ \centering
    \includegraphics[width=0.9\columnwidth]{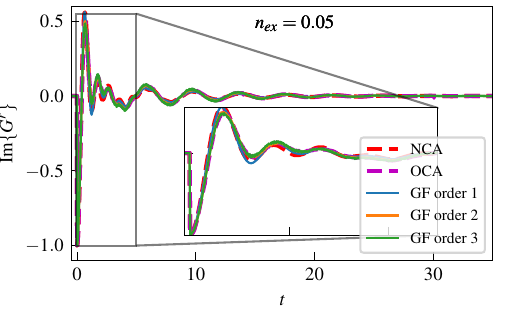}\\
    \caption{ 
            Convergence of the inchworm QMC data with respect to hybridization order. Analyzed is the last DMFT iteration for $n_{\rm {ex}}=0.05$, where the different orders in the hybridization expansion are shown explicitly.
            (a) Restricted propagator.
            (b) Retarded GF. For convenience, the plots also show what an NCA and OCA calculation would yield, given that the exact result obtained within the DMFT scheme is used as their input.
            }
    \label{fig:convergence_inch}
\end{figure}

The inchworm QMC method is based on the hybridization expansion in the coupling between the impurity and its environment, whereby the inchworm approach is an effective resummation scheme that exploits the causal structure of the hybridization expansion.
Consequently, accurate results require convergence in the hybridization order considered.
Fig.~\ref{fig:convergence_inch} 
investigates the convergence of all relevant quantities with respect to the hybridization order for the last DMFT iteration for $n_{\rm {ex}}=0.05$, which pertains data presented in Fig.~2 of the main manuscript.
Fig.~\ref{fig:convergence_inch}(a) shows the restricted propagators, 
$\Phi^\alpha(t) = \mathrm{Tr}_\mathrm{B} \left\lbrace \rho_\mathrm{B} \bra{\alpha} e^{iH t} \ket{\alpha} \right\rbrace$, which is the central quantity for the formulation of the inchworm QMC method. In this notation, $\mathrm{Tr}_\mathrm{B}\lbrace\dots\rbrace$ denotes the trace over the bath degrees of freedom, $\rho_\mathrm{B}$ is the density matrix of the bath, and $\alpha$ is a state of the impurity subsystem. For details we refer to Ref.~\cite{Erpenbeck_Quantum_2023}.
Based on the data presented in Fig.~\ref{fig:convergence_inch}(a), we find that the restricted propagators are essentially converged at second order in the hybridization expansion.

Fig.~\ref{fig:convergence_inch}(b)
shows the retarded GF for hybridization orders $1$--$3$ as calculated from the converged restricted propagators $\Phi^\alpha(t)$. As increasing the order used for calculating the GF also increases the Monte Carlo noise of the calculation, we deem going beyond third order unfeasible for the purpose of this work. 
For convenience, we also show what an NCA and an OCA result at this very DMFT iteration would look like, if the same DMFT hybridization function is used. 

\begin{figure}[h!]
    \centering
    \includegraphics[width=0.9\columnwidth]{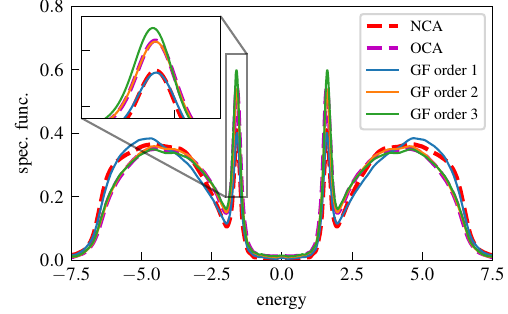}\\
    \caption{ 
            Spectral functions for the Green's functions in Fig.~\ref{fig:convergence_inch}
            The plot also shows what an NCA and OCA calculation would yield, given that the exact result obtained within the DMFT scheme is used as their input.
            }
    \label{fig:convergence_inch_G}
\end{figure}

The corresponding spectra are given in 
Fig.~\ref{fig:convergence_inch_G}.  
While the lowest order GF exhibits differences from the other orders, our results show that the GF for second and third order only show minor
discrepancies, which are reflected in a small difference in the quasiparticle peak height. 
This gives us confidence that GFs calculated at order three provide good accuracy. 

\end{document}